# Integrated Cell Manipulation System – CMOS/Microfluidic Hybrid

**Hakho Lee,[a,c] Yong Liu[b], Donhee Ham[b] and Robert M. Westervelt[*a,b]**

A new type of microfluidic system for biological cell manipulation, a CMOS/microfluidic hybrid, is demonstrated. The hybrid system starts with a custom-designed CMOS (complementary metal-oxide semiconductor) chip fabricated in a semiconductor foundry using standard integration circuit technology. A microfluidic channel is post-fabricated on top of the CMOS chip to provide biocompatible environment. The motion of individual biological cells that are tagged with magnetic beads is directly controlled by the CMOS chip that generates localized magnetic field patterns using an on-chip array of micro-electromagnets. The speed and the programmability of the CMOS chip further allow for the dynamic reconfiguration of the magnetic fields, substantially increasing the manipulation capability of the hybrid system. The concept of a hybrid system is verified by simultaneously manipulating individual biological cells with microscopic resolution. A new operation protocol that exploits the fast speed of electronics to trap and move a large number of cells with less power consumption is also demonstrated. Combining the advantages of microelectronics and microfluidics, the CMOS/microfluidic hybrid approach presents a new model for a multifunctional lab-on-a-chip for biological and biomedical applications.

## Introduction

Spurred by the significant and rapid advances of microelectronics in the past decades, considerable effort has been directed to miniaturize and integrate analytical devices for biological and biomedical applications.[1-3] One major outcome of such efforts is microfluidic systems that are initially developed by adapting the fabrication technology from microelectronics.[4, 5] Consisting of microfabricated plumbing networks, microfluidic systems can generate and control small volumes of fluids, which makes it possible to achieve high throughput while reducing the consumption of reagents or biological samples. Macroscopic experiment setups even can be packaged into compact forms as envisioned in a lab-on-a-chip.[6, 7] Microfluidic systems have been used in various applications including cell sorting,[8] DNA amplifications and separations, [9, 10] and chemical synthesis.[11]

Most of microfluidic systems have fixed structures that are optimized to perform predetermined functions, which is desirable for high-fidelity operations. With their reliance on physical barriers for operation, however, these systems have little flexibility to meet on-demand needs that may arise during experiments. Providing additional functionalities can further incur whole new design and fabrication processes as the structure of the devices has to be modified. To address these issues, multifunctional microfluidic systems based on elaborated fluidic networks, for example, multiplexed channels[12] and modular microfluidic breadboards,[13] have been demonstrated.

[a] *Department of Physics, Harvard University, Cambridge, MA 02138, USA. E-mail: westervelt@deas.harvard.edu*
[b] *Division of Engineering and Applied Sciences, Harvard University, Cambridge, MA 02138, USA.*
[c] *Center for Molecular Imaging Research, Massachusetts General Hospital, Harvard Medical School, Charlestown, MA 02130, USA.*

We have presented a new concept to expand the functionality of microfluidic systems, especially the transport capabilities, by utilizing integrated circuit (IC) chips as an active, versatile actuator inside microfluidic channels.[14] In this approach, IC chips are employed to generate local electromagnetic fields that exert force on target objects in fluids. The fields can be dynamically changed by programming the IC chip to perform many different transport operations. To validate the concept, we have demonstrated the manipulation of magnetic beads using IC/microfluidic chips.[15]

In this paper, we further extend the proposed concept to biological and biomedical applications. Specifically, we report the implementation and the demonstration of a CMOS/microfluidic hybrid system that is designed to manipulate individual biological cells with emphasis on sustaining biocompatible environments during the operation. The hybrid system consists of a custom-designed CMOS (complementary metal-oxide semiconductor) chip and microfluidic channels fabricated on top. The CMOS chip, with micro-electromagnets[16-18] implemented on its surface, directly controls the motion of individual biological cells by generating local magnetic fields. Biological cells are labeled with magnetic beads to be attracted to magnetic fields.[19, 20] The microfluidic channels provide a pathway to introduce cells to the chip, and maintain biocompatible environments. To further enhance the biocompatibility, on-chip temperature sensors are integrated to monitor temperature during experiments. The hybrid system can manipulate individual biological cells with microscopic resolution and in parallel fashion. A protocol that exploits the programmability and the speed of the CMOS chip is also developed to facilitate the chip operation and to bolster biocompatibility.

The unique combination of microelectronics and microfluidics in the CMOS/microfluidic hybrid system presents a new model of a multifunctional microfluidic system for biological and biomedical applications. With CMOS chips functioning as an active component, a hybrid system can perform different, customized tasks just by reprogramming the instructions to the chip, not changing the physical structure of the device. Thus, the hybrid system can serve as a universal platform for a lab-on-a-chip. Furthermore, by integrating control electronics in the same CMOS chip, a hybrid system can readily provide, for example, parallel operations, multiplexing, and user-friendly interfaces that are now common features in microelectronics. Well-established, standard integrated circuit technology also enables the mass production of the chip, which will ensures the consistency in device performances and considerably bring down the cost of the device.

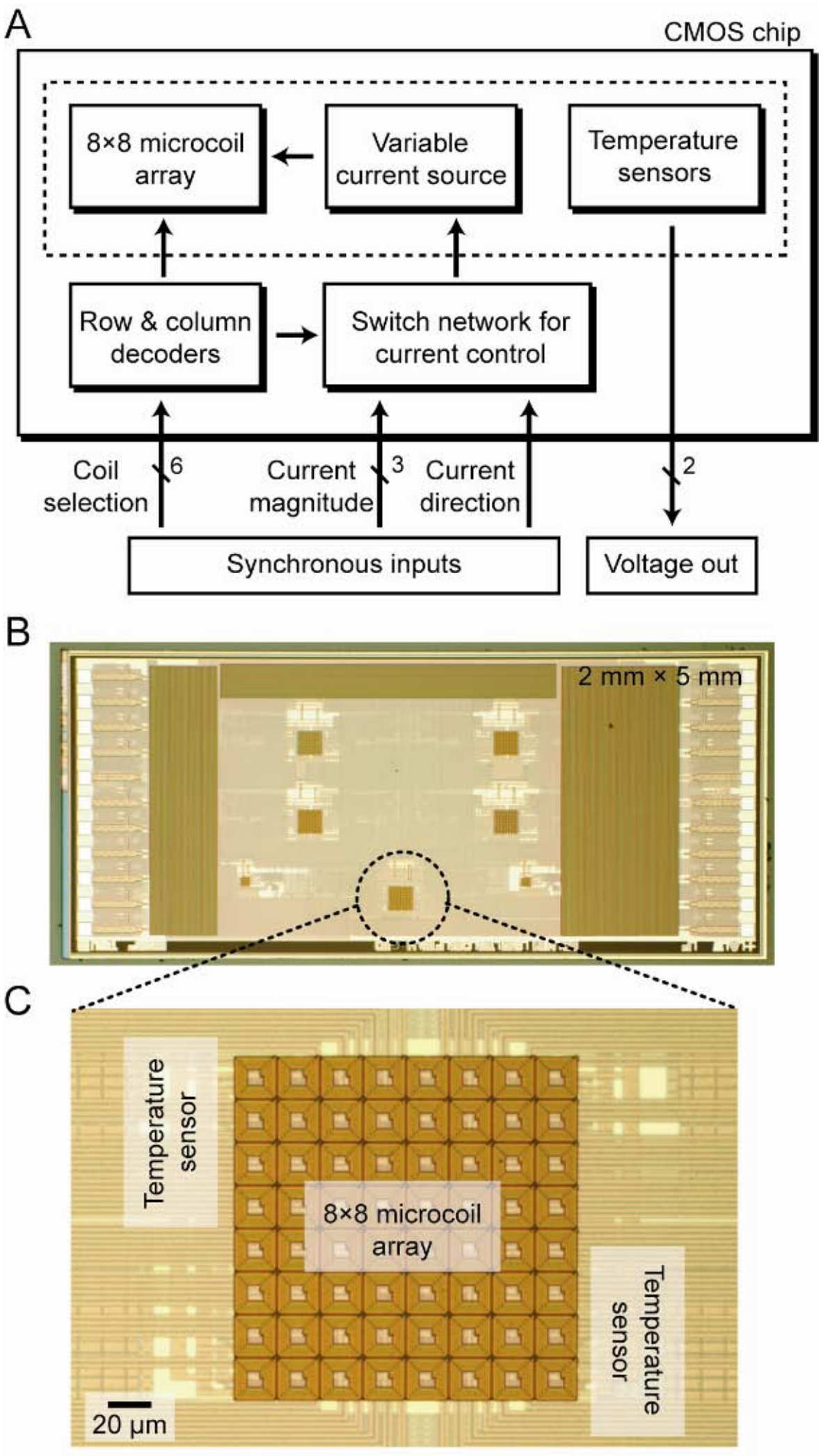

**Fig. 1** Block diagram and micrographs of the CMOS chip for cell manipulation. (A) The CMOS chip contains both analog and digital circuits. Analog circuits (inside the dotted box) include 8×8 microcoil arrays, a current source to the arrays, and temperature sensors. Digital logic circuits, decoders and switch networks control the array by choosing a specific microcoil and setting the direction and the magnitude of the current in the selected microcoil. Digital circuits considerably simplify the interface to the CMOS chip; only ten synchronous inputs are required to individually control 64 microcoils. The temperature sensors output voltages proportional to local temperature near the array. (B) Die micrograph of the CMOS chip. The chip has the footprint of 2×5 mm$^2$, and contains five microcoil arrays. Only one microcoil array is activated during operation. (C) Close-up photo of an 8×8 microcoil array. The outer diameter of a microcoil and the center-to-center pitch between two adjacent microcoils are 20 μm and 21 μm, respectively. Temperature sensors, implemented on the substrate level, are placed near the array.

## CMOS IC Architecture

The CMOS chip for cell manipulation is designed to generate multiple, localized magnetic field peaks on its surface by using an array of surface microcoils. Magnetic manipulation is chosen for two major reasons; 1) magnetic fields are biocompatible and 2) the fields are not influenced by the medium, ensuring high selectivity of magnetically-tagged cells. Generating multiple magnetic peaks is desirable to control the positions of many individual cells in parallel, which can be used, for example, to study intercellular communications and to assemble artificial cellular structures.

Figure 1 shows the functional diagram and the micrograph of the implemented CMOS chip. The chip is fabricated in standard CMOS technology (0.18-μm process of Taiwan Semiconductor Manufacturing Company, Taiwan). The chip contains both analog and digital circuits. The analog circuits include 8×8 microcoil arrays that generate magnetic field patterns for cell manipulation, a variable 8-step current source for the arrays, and on-chip temperature sensors to measure local temperature of the chip. As the digital parts, implemented are row/column decoders to choose a specific microcoil in an array and FET (field-effect transistor) switch networks to control the direction and the magnitude of the electrical current in the selected microcoil. The digital circuits facilitate the operation of the chip by reducing the number of external inputs for chip control. In total, ten synchronous external inputs are required to operate an 8×8 microcoil array; six signals to address individual microcoils, three to set the magnitude of the electrical current in a microcoil, and one to set the direction of the current.

Figure 1B shows the die micrograph of the CMOS chip. The lateral size of the chip is 2 mm × 5 mm, and the operation voltage is $V_{dd}$ = 3.3 V. The chip contains five microcoil arrays with slightly different structure; only one array is enabled during experiments. Figure 1C shows the close-up of one of the microcoil arrays. The array consists of 64 identical microcoils arranged in 8 rows and 8 columns. The outer diameter of the microcoil is 20 μm and the center-to-center distance between two adjacent coils is 21 μm. To accurately monitor the temperature during the cell manipulation, on-chip temperature sensors are placed in the vicinity of the microcoil array.

### Microcoil design.

The goal in the microcoil design is twofold; 1) to increase the trapping accuracy, a single peak in the magnetic field magnitude should be generated at the center of a microcoil on the chip surface, and 2) the field magnitude should be maximized to exert more trapping force on the manipulation target. To meet the design goal, following guidelines are applied when laying out the microcoil structure. First, the outer diameter of the microcoil is chosen to be similar to the size of biological cells to be manipulated. This condition facilitates a single cell trapping by each microcoil. Once the diameter is determined, the number of turns in a microcoil is varied to generate a single magnetic field peak with the maximum magnitude on the chip surface. The magnetic field patterns are calculated using finite element simulation software (MAXWELL 3D, Ansoft, PA). Finally, interconnections between the microcoil and other circuits are made in the metal layer farthest away from the chip surface to minimize stray magnetic fields.

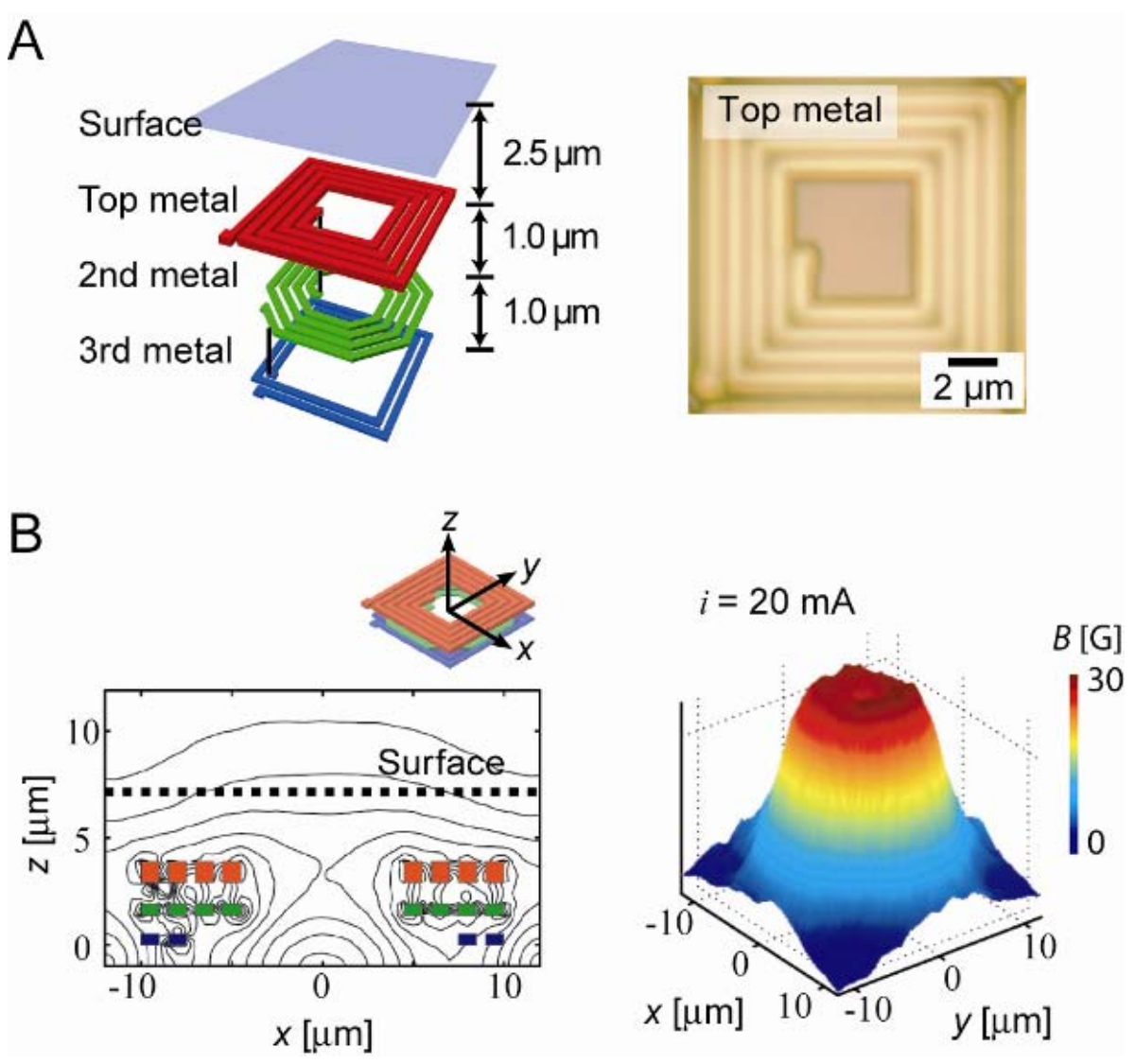

**Fig. 2** Design and magnetic field profiles of a microcoil. The microcoil is designed to generate a single peak in the magnetic field magnitude at its center. (A) The structure of a microcoil. Three planar coils, with each coil in a different metal layer, are connected by vias to form a microcoil. The micrograph shows the planar coil in the top metal layer. The metal lines have the width of 1 μm, and the separation between metal lines is 0.5 μm. Strong magnetic fields can be generated with less current by having a multiple- turn planar coil in the top metal layer. The primary function of the planar coils in the second and the third metal layers is to shape a single peak in the field magnitude. (B) Calculated magnetic field profiles across the microcoil and on the surface of the CMOS chip. A single peak is formed on the chip surface at the center of the microcoil.

Figure 2A shows the micrograph and the structure of the microcoil implemented in the CMOS chip. Three metal layers (the top, the second, and the third counting from the chip surface) are used with each metal layer containing a planar coil. To form a microcoil, the planar coils are connected by vias in series. Adjacent metal layers are separated by an 1 μm thick $SiO_2$ layer, and the top surface of the chip is passivated with a 2.5 μm thick polyimide. Figure 2B shows the calculated magnetic field profiles across the microcoil and on the chip surface. Because most of contribution to the field magnitude comes from the planar coil in the top metal layer, the planar coil in the top metal layer has multiple turns to generate stronger magnetic fields with less current. Planar coils in the second and the third metal layers are primarily used to shape a single peak at the center of the microcoil on the chip surface.

### Digital control circuit

To provide an independent control on the electrical current in each microcoil, the CMOS chip integrates multiplexing circuitry that includes row/column decoders, logic gates and switching networks. Figure 3A shows the schematic of the control circuit. The components inside the frame are repeated for each microcoil; the decoders and the on-chip current source are shared in a microcoil array. When the FET switch ($S_c$) is activated by the decoder outputs, the microcoil is connected to the on-chip current source. Note that only one microcoil in an array can access the current source at any given moment. The current source can sink up to 20 mA in the increment of 2.5 mA, providing an 8-step control in the current magnitude. The direction of the current is set by two FET switches ($S_{D1}$ and $S_{D2}$) that are gated by the external direction signal and the decoder signals. Together with the 8-step current source, the direction control realizes a bipolar, variable current source to each microcoil.

To generate many magnetic peaks for multiple cell trapping, the common current source is shared in time domain. For example, to create $N^2$ magnetic peaks using an $N\times N$ microcoil array, a current pulse with the duty cycle of $1/N^2$ is sequentially applied to each microcoil, which generates a "flashing" magnetic peak scanning through the array. If the period of the current pulse is smaller than the escape time of a trapped cell due to Brownian motion, the flashing magnetic peak by the current pulse can be averaged as a stationary peak, effectively holding the cell at the same location. Because the estimated escape time of the cells used in this report is > 10 sec, whereas the minimum pulse period that CMOS chip can handle is << 1 μsec, the current source can be easily shared in time domain to imitate the simultaneous generation of multiple magnetic peaks. Note that a method in the same principle is employed in optical tweezers to create multiple optical traps from a single light source.[21]

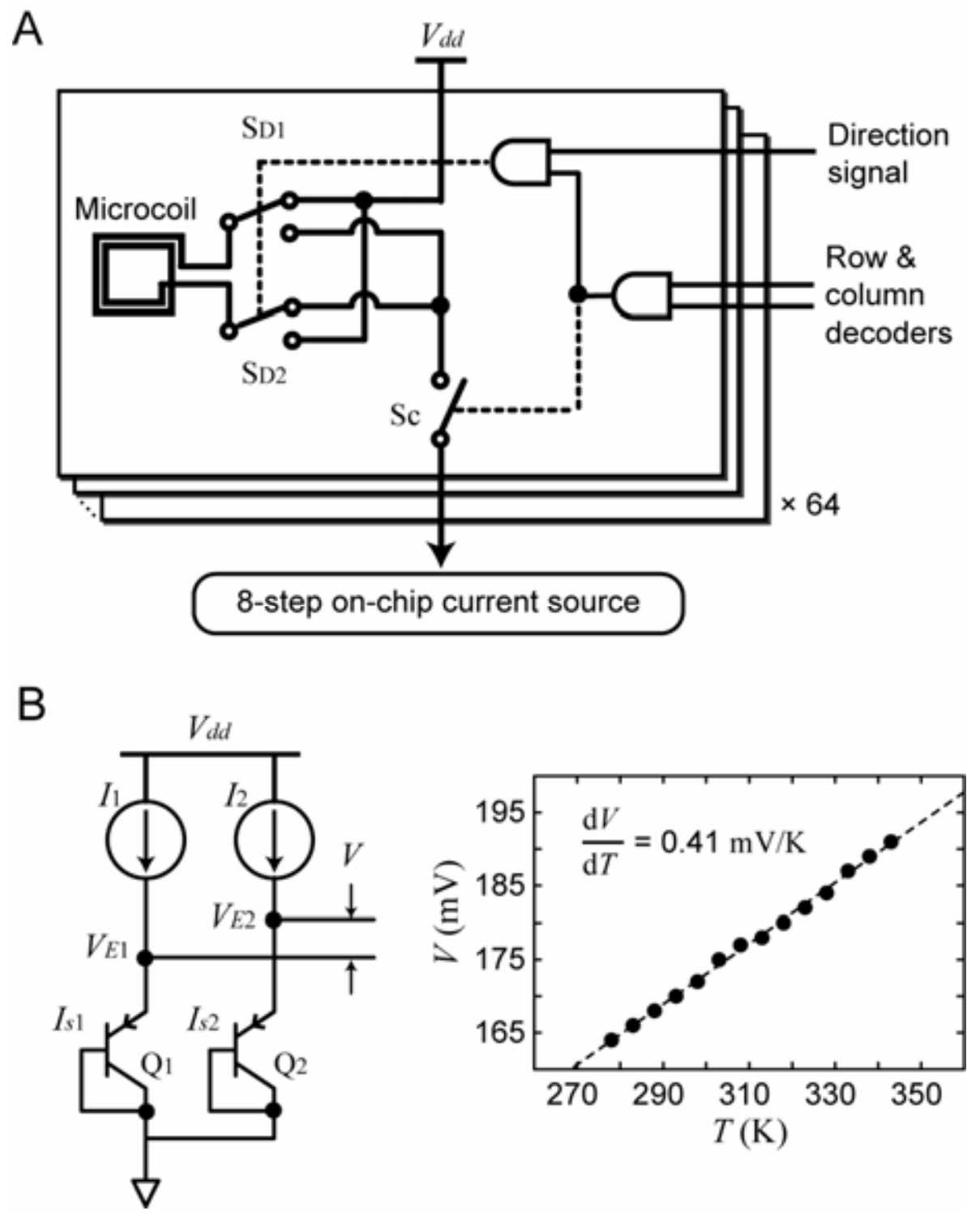

**Fig. 3** (A) Circuit schematic of a microcoil. To enable multiplexing, each microcoil circuit has three FET (field effect transistor) switches and two AND logic gates. The microcoil is connected to the on-chip current source through the switch Sc that is gated by the AND-logic of row and column decoder signals. The current direction is set by the $S_{D1}$ and $S_{D2}$ switches that are controlled by the external direction signal. The on-chip current source provides an 8-step control in the current magnitude with the maximum output of 20 mA. The circuitry inside the frame is repeated for all 64 microcoils; the current source and the decoders are shared in a given microcoil array. (B) Circuit schematic and measured characteristics of the on-chip temperature sensor. The sensor uses two PNP transistors in a diode connection. The emitter voltages ($V_{E1}$ and $V_{E2}$) of the transistors ($Q_1$ and $Q_2$) are directly proportional to absolute temperature. By measuring the difference $V$ between $V_{E1}$ and $V_{E2}$, errors coming from the drift of transistors can be compensated. Measured output $V$ from the temperature sensor as a function of temperature $T$ shows a good linear relationship with $dV/dT$ = 0.41 mV/K.

The time-sharing of a current source, together with the multiplexing scheme, brings several advantages, especially when the number of microcoils in an array is large. First, microcoils can still be individually controlled by a small number of external inputs, simplifying the interface to the CMOS chip. Second, because only one microcoil is activated at any given moment, a common current source is sufficient to operate the whole array, making it easier to design and implement the current source. Third, the power consumption in the microcoil array remains nearly constant regardless of the number of microcoils, which minimizes the heat generation during cell manipulation and further enhances biocompatibility.

### Temperature sensors

Regulating the temperature of the IC during the cell manipulation is important to minimize the thermal damage to cells. To monitor local temperature on the chip during cell manipulation, on-chip temperature sensors are implemented in the vicinity of microcoil arrays. Figure 3B shows the schematic and the measured characteristics of the on-chip temperature sensor. The sensor is implemented using parasitic PNP bipolar transistors available in a standard CMOS process.[22] With a constant current $I_1$, the emitter voltage $V_{E1}$ in the transistor $Q_1$ is directly proportional to the absolute temperature $T$, $V_{E1} = (k_B T/q)\log(I_1/I_{s1})$, where $q$ is the electron charge, $k_B$ is the Boltzmann constant, and $I_{s1}$ is the leakage current in $Q_1$. Placing another transistor $Q_2$ with an additional current source and by measuring the emitter voltage differences $V = V_{E1}$-$V_{E2}$, errors coming from the variations of transistors can be compensated. In the actual chip design, $I_2 = 0.25I_1$ and $I_{s2} = 5I_{s1}$, which gives the relationship $\mathrm{d}V/\mathrm{d}T$ = 0.40 mV/K, where $I_2$ is the current from the additional current source to $Q_2$ and $I_{s2}$ is the leakage current in $Q_2$. The measured sensor output ($\mathrm{d}V/\mathrm{d}T$ = 0.41 mV/K) agrees well with the expected linear dependency.

## CMOS/Microfluidic Hybrid System

After receiving the CMOS chips from the foundry, microfluidic channels are post-fabricated on top of the IC chips to complete the hybrid system. The microfluidic channels provide a pathway to introduce biological cells to the chip surface, and maintain biocompatible environment during cell manipulation.

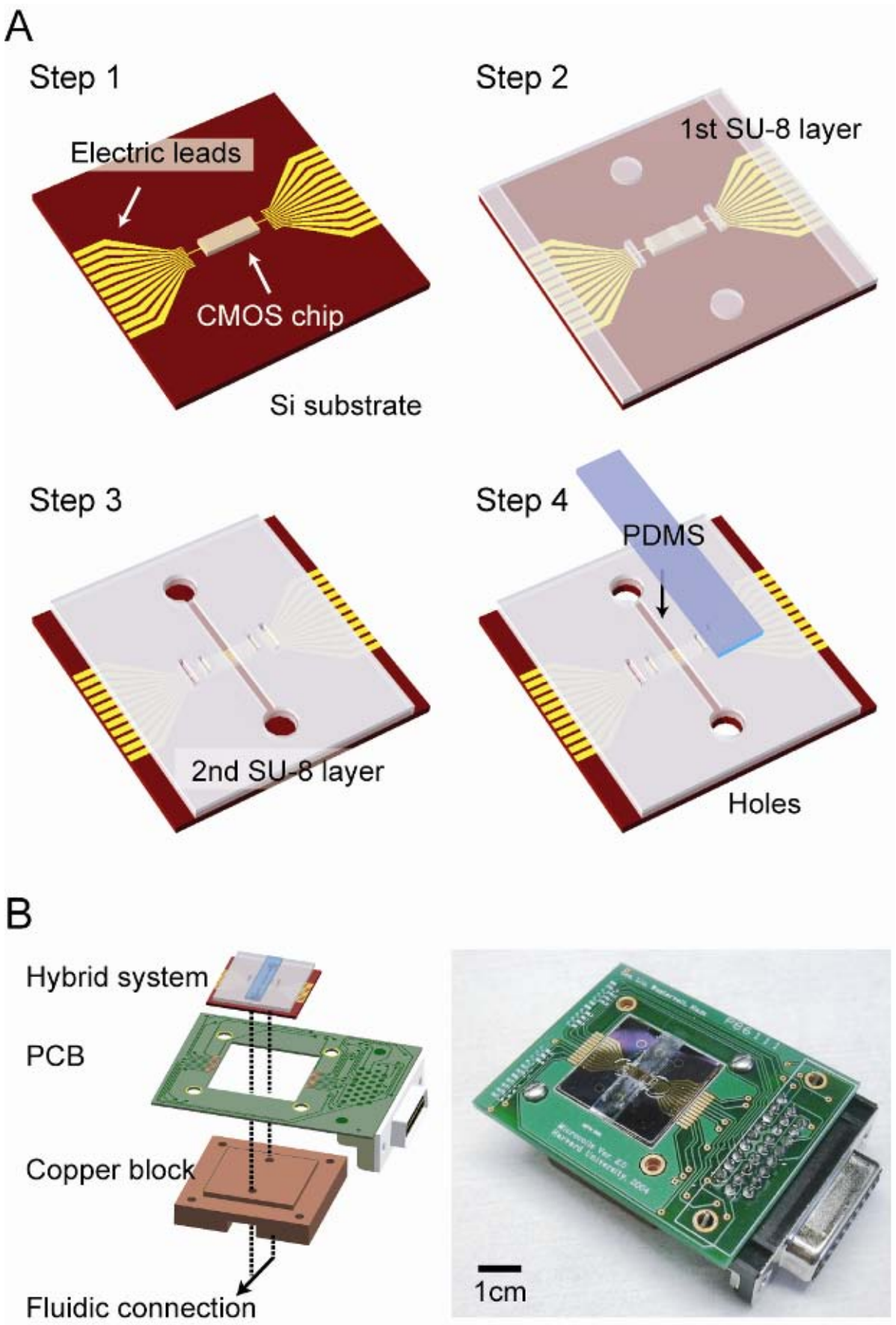

**Fig. 4** (A) Fabrication process of microfluidic channels. [Step 1] The CMOS chip is glued to a $Si/SiO_2$ substrate that has lithographically patterned electrodes. [Step 2] The first layer of SU-8 is spin-coated and patterned, but is not developed. This layer provides a leveled surface for the subsequent microfluidic channel fabrication. [Step 3] The second SU-8 layer is spin-coated and patterned, and both SU-8 layers are developed. The layers are then hard-cured on a hotplate. [Step 4] Holes are drilled on the substrate for fluidic connections and the microfluidic channel is sealed with a cured PDMS (polydimethylsiloxane) layer. (B) Packaging of the CMOS/microfluidic hybrid system. The assembled hybrid system is mounted on a copper block, and is electrically connected to a PCB (print circuit board) via wire bonding. Fluidic connections are made through the backside of the hybrid system, freeing the front side of the system from fluidic tubing components. The packaged system is then mounted on a microscope stage cooled by a thermoelectric cooler.

### Fabrication of microfluidic channel

Because the bare CMOS chip has a small footprint (2 mm x 5 mm), the chip is attached to a bigger $Si/SiO_2$ substrate for the subsequent fabrication. The fluidic structure is fabricated using the double layer technique to form microfluidic channels with a uniform depth,[23] which minimizes the accumulation of cells near the CMOS chip edges during cell injection.

Figure 4A illustrates the major steps in the post-fabrication process. First, the chip is glued to a $Si/SiO_2$ substrate with electrical lead patterns [Step 1]. Subsequently, a layer of SU-8 is spin-coated and exposed to UV light to pattern fluidic ports and openings for electrical connections. However, the resist is not developed [Step 2]. The thickness of the first SU-8 layer is comparable to that of the CMOS chip (~ 350 μm) to provide a leveled surface. In the next step, a second layer of SU-8 is spin-coated and patterned for a microfluidic channel, and then both SU-8 layers are developed [Step 3]. The resulting microfluidic channel has relatively uniform depth (~ 120 µm in the CMOS chip area and 100 µm elsewhere). After hard-curing the SU-8 layers, holes are drilled on the Si/$SiO_2$ substrate to form fluidic inlets, and a layer of thin PDMS (polydimethylsiloxane) is placed on top of the SU-8 structure to seal the microfluidic channels [Step 4].

**System packaging and operation**

The assembled hybrid system is packaged for operation as shown in Fig. 4B. The whole system is mounted on a copper stage for cooling, and electrically connected to a printed circuit board (PCB) via wire bonding. Fluidic connections are made from the backside of the hybrid system, which frees the front side from fluidic components. This fluidic connection scheme allows the hybrid system to have a low profile, making it easier to mount the system under a microscope.

The packaged hybrid system is mounted on a microscope stage that is cooled by a thermoelectric cooler unit (TEC 2000, ThorLabs Inc., NJ). The manipulation process is observed by a microscope equipped with a CCD camera. To operate the CMOS chip in the hybrid system, a microcomputer board with a microcontroller (ATmega32, Atmel Corporation) is custom-made. The microcomputer generates all external control signals to the CMOS chip, reads out the on-chip temperature sensors.

**Cell preparation**

As a representative manipulation target, bovine capillary endothelial (BCE) cells are used. To impart magnetic moment to the cells, peptide-coated magnetic beads are co-cultured with BCE cells, which leads to the uptake of the beads by the cells through endocytosis. Specifically, magnetic beads of diameter 250 nm and with –COOH functional surface group (PCM-250, Kisker, Germany) are coated with GRGDSP peptide sequence (#22946, AnaSpec, CA) using water-soluble carbodiimide chemistry.[24-26] Bovine capillary endothelial (BCE) cells (CRL-8659, ATCC, VA) are cultured in the presence of the magnetic beads (5 µg/ml) for 24 hours. For the cell manipulation experiments, BCE cells are trypsinzied and suspended in PBS. Before introducing the cells, the CMOS chip surface is coated with bovine serum albumin (BSA) by filling the microfluidic channel with BSA solution (5% w/v) to suppress unspecific cell binding to the chip surface. Magnetically-tagged BCE cells are then introduced onto the microcoil array through the microfluidic channel.

## Results and Discussion

Experiments with magnetically-tagged cells are performed to demonstrate the several important capabilities of the hybrid system: 1) the control of individual cells with microscopic resolution, 2) the dynamic reconfiguration of magnetic fields patterns for versatile cell manipulation, and 3) the stable trapping and transport of multiple cells while sharing a single current source in time domain.

Figure 5 shows the trapping and the transport of a single BCE cell using the hybrid system. By applying a current pulse of 20 mA to a microcoil, a magnetic field peak with the magnitude of 30 G is generated, trapping a single cell at the center of the microcoil [Fig. 5A]. The period of the current pulse is 100 msec with the duty cycle of 50%. Because the diffusion of the cell due to Brownian motion is negligible while the current is off, the cell can be effectively trapped at the same position. To move the cell, the current pulse is applied to an adjacent microcoil, which results in the hopping of the cell from one coil center to another center. The same procedure is repeated to perform a complete round trip of the cell [Fig. 5B].

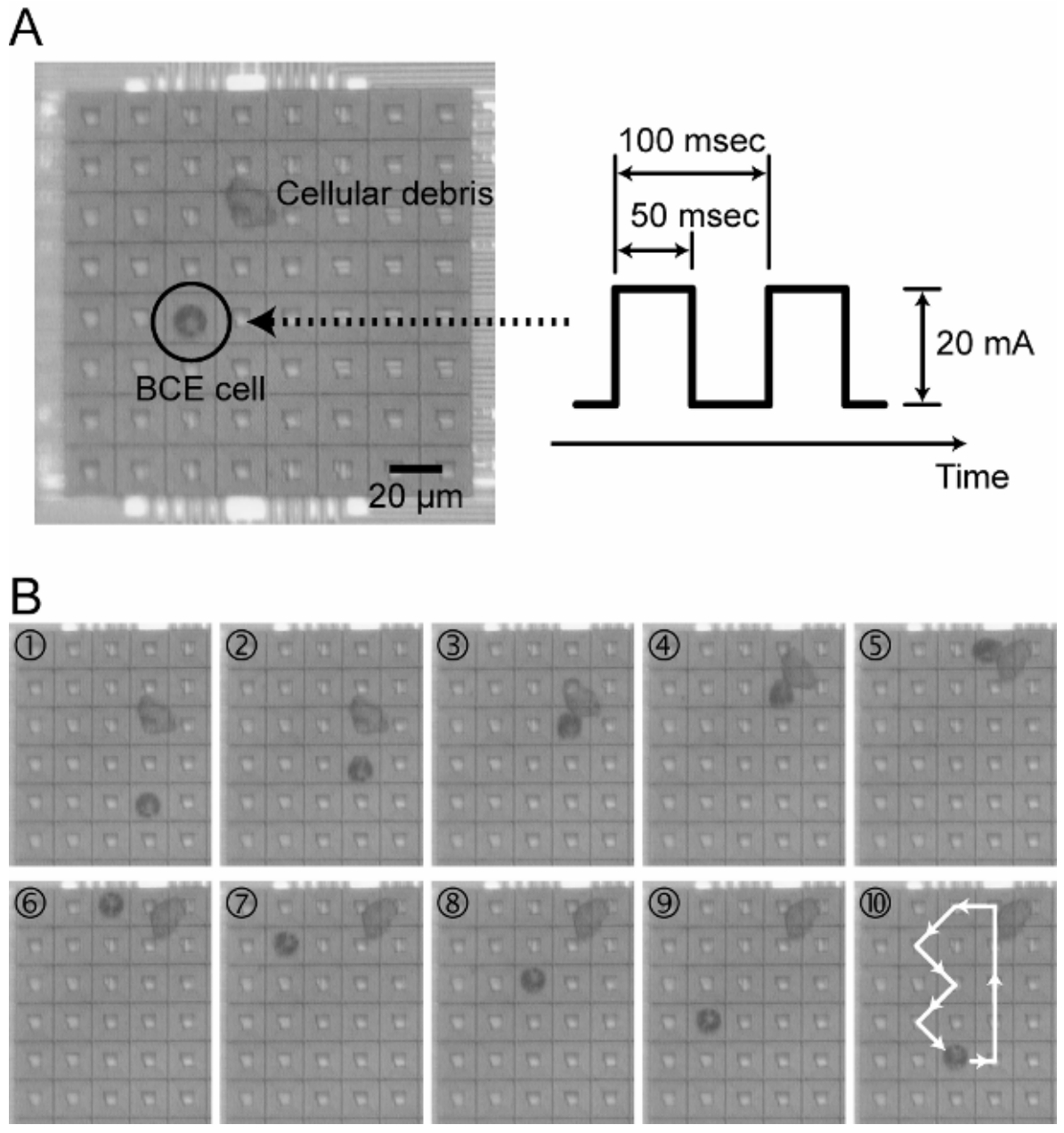

**Fig. 5** Manipulation of a single BCE (bovine capillary endothelial) cell using the hybrid system. BCE cells are magnetically tagged; magnetic beads of diameter 250 nm are taken up by the cells through endocytosis. (A) A single BCE cell is trapped at the center of a microcoil. A current with the magnitude of 20 mA is pulsed to the microcoil. The period and the duty cycle of the pulse are 100 msec and 50%, respectively. (B) The trapped cell is moved over the microcoil array to make a round trip. The same current pulse in (A) is applied to the microcoil adjacent to the cell, which results in the hopping of the cell from one coil center to another center. During the transport, the cellular debris is pushed away.

Besides creating a single magnetic peak at each microcoil, one can generate more versatile magnetic field patterns by adjusting the magnitude and the direction of currents in microcoils. As an example, Fig. 6 shows how a trapped cell can be positioned at locations other than the centers of microcoils, increasing the spatial resolution of cell trapping. By controlling the current distribution in two adjacent microcoils, a single magnetic peak can be moved in the steps less than the center-to-center pitch of the microcoils [Fig. 6A]. Note that during the peak movements, the magnitude of the peak remains nearly constant, but the direction of the magnetic field is changing. Figure 6B shows the step-wise transport of a BCE trapped cell over two neighboring microcoils. To mimic the magnetic field patterns shown in Fig. 6A, currents of appropriate magnitude and polarity are pulsed to each coil alternatively with the period of 20 msec and the duty cycle of 50%. The trapped cell is observed to roll on the chip surface during the transport, as the direction of the magnetic field is changing at each peak position.

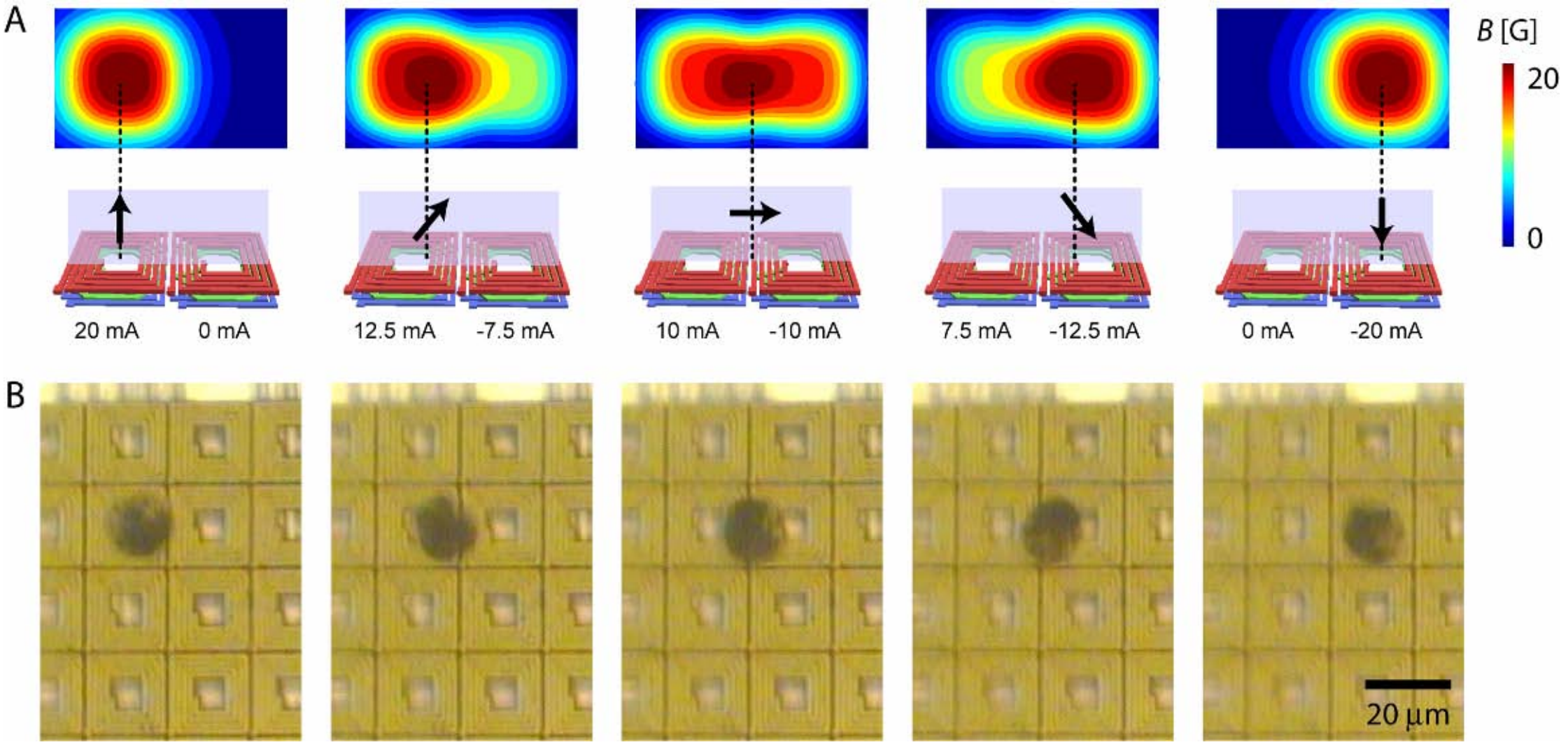

**Fig. 6** By controlling currents in the microcoils, many different magnetic field patterns can be generated to provide more flexibility in cell manipulation. As an example, a single peak in the magnetic field magnitude is created and moved in steps over two neighboring microcoils. (A) Calculated magnetic field patterns with current distribution in microcoils. The magnitude of the field peak remains nearly the same, but the direction of the field rotates. The arrows indicate the magnetic field direction in the plane normal to the chip surface. (B) A single BCE cell is trapped and moved using the field pattern in (A). To generate the fields, current pulses with appropriate values and directions are alternatively applied to two microcoils. The pulse period is 20 msec and the duty cycle is 50%. The cell is observed rolling on the chip surface during the transport due the change of the field direction.

To verify the time-sharing of a single current source, multiple cells are simultaneously manipulated as shown in Fig. 7. Initially, three BCE cells are trapped by sequentially pulsing currents to three microcoils. The period, the duty cycle, and the height of the current pulse are 30 msec, 33%, and 20 mA, respectively. With a single cell held still, the rest of the cells are moved independently by activating appropriate microcoils with the same pulse profile, and finally all cells are arranged next to each other. No cell is lost during this operation, confirming the time-sharing method as an effective way to generate multiple magnetic traps.

The capability of the hybrid system to individually control many cells in parallel can be used to conduct many new types of experiments with single-cell level precision. For example, cells of different types can be brought together to study intercellular communications[27]; biological tissues with an artificial cell composition can be assembled by bringing cells, one-by-one, in a controlled manner.[28, 29]

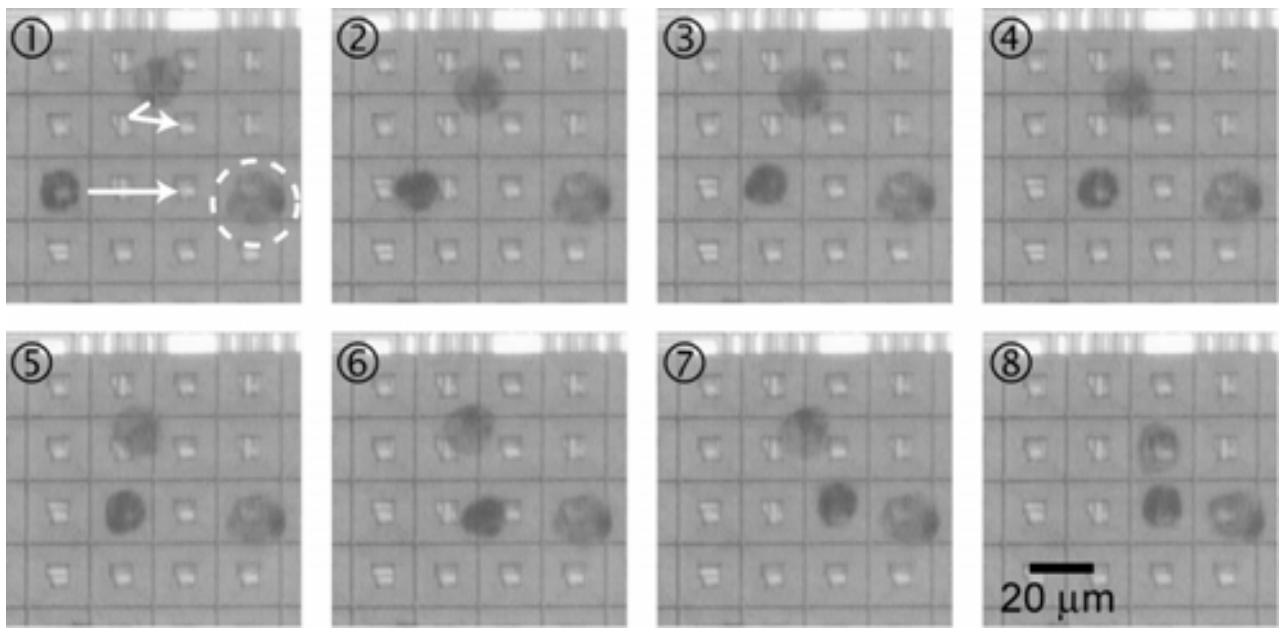

**Fig. 7** Manipulation of multiple cells. Three BCE cells are trapped and independently controlled to sit next to each other; the cell in the dotted circle is held still, the rest of the cells are moved. The on-chip current source is sequentially shared in time domain among three microcoils; the current of 20 mA is on for 10 msec and off for 20 msec in each microcoil. Because the diffusion of trapped cells is negligible during the off-time of currents, multiple magnetic traps can be effectively created simultaneously. The time-sharing method provides a practical solution to handling large microcoil arrays.

## Conclusions

The CMOS/microfluidic hybrid system presents a new direction for multifunctional microfluidic systems in biological and medical applications. Exploiting the speed and the programmability of the IC chips, the hybrid system brings flexibility inside microfluidic channels. Mature semiconductor technology enables the integration of many functional electronic circuits in a single chip, so one can make self-contained hybrid systems with uniform performance from chip to chip. Because the integrated circuits in hybrid systems can be manufactured as conventional ICs in a semiconductor foundry, their cost can be drastically reduced like consumer electronics –a hybrid system can be a cheap and disposable device.

As a proof-of-concept, we have implemented a hybrid prototype that uses a CMOS chip to move biological cells by generating spatially patterned magnetic fields. The motion of individual cells has been controlled at microscopic resolution using microcoil arrays; on-chip temperature sensors and integrated digital circuits facilitate the chip operation as well as maintaining a biocompatible environment. This hybrid system provides a novel way to conduct experiments with single-cell precision in a reproducible fashion, which will be very helpful in understanding intercellular processes.

The approach reported here can be used to implement other types of IC/microfluidic hybrid systems that are dedicated to specific biological and medical applications. For example, hybrid IC/Microfluidic chips can be used to sort cells based on dielectrophoresis[30], to monitor and stimulate electrogenic cells,[31, 32] and to detect DNA hybridization[33]. Bringing advanced microelectronics and microfluidics together, the hybrid system will be a powerful lab-on-a-chip platform.

## Acknowledgement

The authors thank L. DeVito and S. Feindt of Analog Devices for their support in chip fabrication, and E. Alsberg and D. Ingber at Harvard Medical School for helpful discussion. EM field solvers were donated by AnSoft and Sonnet. This work is supported by Nanoscale Science and Engineering Center (NSEC) at Harvard under NSF grant PHY-0117795 and IBM Faculty Partnership Award.